%%%%%%%%%%%%%%%%%%%%%%%%%%%%%%%%%%%%%%%%%%%%%%%%%%%%%%%%%%%%%%%%%%%%%%
%%%%%%%%%%%%%%%%%%%%%%%%%%%%%%%%%%%%%%%%%%%%%%%%%%%%%%%%%%%%%%%%%%%%%%
%%%                                                                %%%
%%%         .         .           ..           .         .         %%%
%%%             Parrondo games as lattice gas automata             %%%
%%%                                                                %%%
%%%               David A. Meyer and Heather Blumer                %%%
%%%                                                                %%%
%%%                                                                %%%
%%%%%%%%%%%%%%%%%%%%%%%%%%%%%%%%%%%%%%%%%%%%%%%%%%%%%%%%%%%%%%%%%%%%%%
%%%%%%%%%%%%%%%%%%%%%%%%%%%%%%%%%%%%%%%%%%%%%%%%%%%%%%%%%%%%%%%%%%%%%%
%%%                                                                %%%
%%%                                                                %%%
%%%                       Typesetting notes                        %%%
%%%                                                                %%%
%%%                                                                %%%
%%% Plain TeX with a few simple macros at the beginning.           %%%
%%%                                                                %%%
%%% If AMS fonts are not loaded, comment the next line and         %%%
%%%  uncomment the following one.                                  %%%
\font\bbb=msbm10 \font\bbs=msbm7                                   %%%
%\def\bbb{\bf} \def\bbs{\bf}                                       %%%
%%%                                                                %%%
%%%                                                                %%%
%%%%%%%%%%%%%%%%%%%%%%%%%%%%%%%%%%%%%%%%%%%%%%%%%%%%%%%%%%%%%%%%%%%%%%
%%%%%%%%%%%%%%%%%%%%%%%%%%%%%%%%%%%%%%%%%%%%%%%%%%%%%%%%%%%%%%%%%%%%%%

\overfullrule=0pt

\magnification=1200

\def\C{\hbox{\bbb C}}  
  
\def\R{\hbox{\bbb R}}  
\def\Z{\hbox{\bbb Z}} \def\sZ{\hbox{\bbs Z}}

\def\AER{{\sl Amer.\ Econom.\ Rev.}}

\def\AJP{{\sl Amer.\ J. Phys.}}

\def\AdP{{\sl Ann.\ Physik}}

\def\CJP{{\sl Canadian J. Phys.}}

\def\CPAM{{\sl Commun.\ Pure Appl.\ Math.}}

\def\IJMPC{{\sl Int.\ J. Mod.\ Phys.\ C}}

\def\INCD{{\sl Il Nuovo Cimento D}}

\def\JMP{{\sl J. Math.\ Phys.}}
\def\JPA{{\sl J. Phys.\ A:  Math.\ Gen.}}

\def\JPSJ{{\sl J. Phys.\ Soc.\ Jpn.}}
\def\JSP{{\sl J. Statist.\ Phys.}}

\def\LNC{{\sl Lett.\ Nuovo Cimento}}

\def\Na{{\sl Nature\/}}

\def\NCA{{\sl Nuovo Cimento A}}

\def\PA{{\sl Physica A\/}}
\def\PB{{\sl Physica B\/}}

\def\PLA{{\sl Phys.\ Lett.\ A\/}}

\def\PLMS{{\sl Proc.\ Lond.\ Math.\ Soc.}}
\def\PM{{\sl Philos.\ Mag.}}

\def\PRA{{\sl Phys.\ Rev.\ A\/}}

\def\PRE{{\sl Phys.\ Rev.\ E\/}}
\def\PRL{{\sl Phys.\ Rev.\ Lett.}}

\def\PZ{{\sl Phys.\ Z.}}
\def\QJMAM{{\sl Quart.\ J. Mech.\ Appl.\ Math.}}
\def\RMJM{{\sl Rocky Mountain J. Math.}}

\def\Sc{{\sl Science\/}}

\def\StSc{{\sl Stat.\ Science}}

\def\ZP{{\sl Z. Physik\/}}

\def\dajm{\hbox{D. A. Meyer}}

\def\brosl{\hbox{B. Hasslacher}}
\def\bd{\hbox{\brosl\ and \dajm}}

\def\ha{\hbox{G. P. Harmer and D. Abbott}}
\def\hpdp{\hbox{J. Hardy, Y. Pomeau and O. de Pazzis}}
\def\hdpp{\hbox{J. Hardy, O. de Pazzis and Y. Pomeau}}

\def\hfb{\hfil\break}

\catcode`@=11
\newskip\ttglue

   \font\ninerm=cmr9    \font\eightrm=cmr8   \font\sixrm=cmr6
  \font\ninebf=cmbx9   \font\eightbf=cmbx8  \font\sixbf=cmbx6
  \font\nineit=cmti9   \font\eightit=cmti8  
  \font\ninesl=cmsl9   \font\eightsl=cmsl8  
  \font\ninemi=cmmi9   \font\eightmi=cmmi8  \font\sixmi=cmmi6

\font\bigten=cmr10 scaled\magstep2 

\def\ninepoint{\def\rm{\fam0\ninerm}%
  \textfont0=\ninerm \scriptfont0=\sixrm
  \textfont1=\ninemi \scriptfont1=\sixmi
  \textfont\itfam=\nineit  \def\it{\fam\itfam\nineit}%
  \textfont\slfam=\ninesl  \def\sl{\fam\slfam\ninesl}%
  \textfont\bffam=\ninebf  \scriptfont\bffam=\sixbf
    \def\bf{\fam\bffam\ninebf}%
  \tt \ttglue=.5em plus.25em minus.15em
  \normalbaselineskip=11pt
  \setbox\strutbox=\hbox{\vrule height8pt depth3pt width0pt}%
  \normalbaselines\rm}

\def\eightpoint{\def\rm{\fam0\eightrm}%
  \textfont0=\eightrm \scriptfont0=\sixrm
  \textfont1=\eightmi \scriptfont1=\sixmi
  \textfont\itfam=\eightit  \def\it{\fam\itfam\eightit}%
  \textfont\slfam=\eightsl  \def\sl{\fam\slfam\eightsl}%
  \textfont\bffam=\eightbf  \scriptfont\bffam=\sixbf
    \def\bf{\fam\bffam\eightbf}%
  \tt \ttglue=.5em plus.25em minus.15em
  \normalbaselineskip=9pt
  \setbox\strutbox=\hbox{\vrule height7pt depth2pt width0pt}%
  \normalbaselines\rm}

\def\sfootnote#1{\edef\@sf{\spacefactor\the\spacefactor}#1\@sf
      \insert\footins\bgroup\eightpoint
      \interlinepenalty100 \let\par=\endgraf
        \leftskip=0pt \rightskip=0pt
        \splittopskip=10pt plus 1pt minus 1pt \floatingpenalty=20000
        \parskip=0pt\smallskip\item{#1}\bgroup\strut\aftergroup\@foot\let\next}
\skip\footins=12pt plus 2pt minus 2pt
\dimen\footins=30pc

\def\ie{{\it i.e.}}
\def\eg{{\it e.g.}}

\def\and{{\eightpoint AND}}

\def\PQpf{PQ P{\eightpoint ENNY} F{\eightpoint LIP}}

\def\FeynmanHibbs{1}
\def\qcaqlg{2}
\def\BoghosianTaylor{3}
\def\qlgaI{4}
\def\qlgaII{5}
\def\Parrondo{6}
\def\HarmerAbbott{7}
\def\qstrat{8}
\def\Abbott{9}
\def\Taylor{10}
\def\DWMF{11}
\def\diffusion{12}
\def\spurious{13}
\def\flashing{14}
\def\Smoluchowski{15}
\def\Feynman{16}
\def\ParrondoEspanol{17}
\def\ADP{18}
\def\electro{19}
\def\optical{20}
\def\qrattheory{21}
\def\qratexp{22}
\def\nogo{23}
\def\fluidflow{24}
\def\Goldstein{25}
\def\Kac{26}
\def\PHA{27}
\def\lgbu{28}
\def\DO{29}
\def\NTC{30}
\def\BenjaminHayden{31}
\def\minority{32}
\def\AAKV{33}
\def\NayakVishwanath{34}
\def\CFG{35}
\def\NielsenChuang{36}
\def\pqc{37}

\input epsf.tex

\dimen0=\hsize \divide\dimen0 by 13 \dimendef\chasm=0
\dimen1=\hsize \advance\dimen1 by -\chasm \dimendef\usewidth=1
\dimen2=\usewidth \divide\dimen2 by 2 \dimendef\halfwidth=2
\dimen3=\usewidth \divide\dimen3 by 3 \dimendef\thirdwidth=3
\dimen4=\hsize \advance\dimen4 by -\halfwidth \dimendef\secondstart=4
\dimen5=\halfwidth \advance\dimen5 by -10pt \dimendef\indenthalfwidth=5
\dimen6=\thirdwidth \multiply\dimen6 by 2 \dimendef\twothirdswidth=6
\dimen7=\twothirdswidth \divide\dimen7 by 4 \dimendef\qttw=7
\dimen8=\qttw \divide\dimen8 by 4 \dimendef\qqttw=8
\dimen9=\qqttw \divide\dimen9 by 4 \dimendef\qqqttw=9
%\dimen10=\chasm \multiply\dimen10 by 10 \dimendef\fivein=10
\dimen10=2.25truein \dimendef\sciencecol=10

\parskip=0pt\parindent=0pt

\line{\hfil 15 February 2001}
%\line{\hfil {\it revised\/} 2 August 2001}
\line{\hfill quant-ph/0110028}
\vfill
\centerline{\bf\bigten PARRONDO GAMES AS}
\smallskip
\centerline{\bf\bigten LATTICE GAS AUTOMATA}
\bigskip\bigskip
\centerline{\bf David A. Meyer$^{*\dagger}$ and 
                Heather Blumer$^{\dagger}$}
\bigskip 
\centerline{\sl {}$^*$Project in Geometry and Physics,
                Department of Mathematics}
\centerline{\sl University of California/San Diego,
                 La Jolla, CA 92093-0112}
\smallskip
\centerline{\sl {}$^{\dagger}$Institute for Physical Sciences,
                 Los Alamos, NM 87544}
\smallskip
\centerline{{\tt dmeyer@chonji.ucsd.edu}, {\tt hblumer@lanl.gov}}
\smallskip

\vfill
\centerline{ABSTRACT}
\bigskip
%--------|---------|---------|---------|---------|---------|---------|
\noindent Parrondo games are coin flipping games with the surprising
property that alternating plays of two losing games can produce a 
winning game.  We show that this phenomenon can be modelled by 
probabilistic lattice gas automata.  Furthermore, motivated by the
recent introduction of {\sl quantum\/} coin flipping games, we show
that quantum lattice gas automata provide an interesting definition
for quantum Parrondo games.
\bigskip\bigskip
%--------|---------|---------|---------|---------|---------|---------|
\noindent 2001 Physics and Astronomy Classification Scheme:
                   02.70.Ns, % Molecular dynamics and particle methods
                   05.40.Fb, % Random walks and Levy flights
                   03.65.Pm, % Relativistic wave equations
                   02.50.Le. % Decision theory and game theory

\noindent 2000 American Mathematical Society Subject Classification:
                   60J60,    % Diffusion processes
                   82C10,    % Quantum dynamics and nonequilibrium 
                             %  statistical mechanics
                   60G50,    % Sums of independent random variables;
                             %  random walks
                   91A15.    % Stochastic games
\smallskip
\global\setbox1=\hbox{Key Words:\enspace}
\parindent=\wd1
\item{Key Words:}  Parrondo games, lattice gas automata, quantum 
                   games, quantum lattice gas automata, correlated 
                   random walk.

\vfill
\hrule width2.0truein
\medskip
\noindent Expanded version of an invited talk presented at the Ninth 
International Conference on Discrete Models in Fluid Mechanics held 
in Santa Fe, NM, 21--24 August 2000.
\eject

\headline{\ninepoint\it Parrondo games as LGA  \hfill Meyer \& Blumer}

\parskip=10pt
\parindent=20pt

\noindent{\bf 0.  Introduction}

%--------|---------|---------|---------|---------|---------|---------|
\noindent The simplest quantum lattice gas automata (QLGA) provide 
discrete models for the $1+1$ dimensional Dirac equation 
[\FeynmanHibbs,\qcaqlg] and the multiparticle Schr\"odinger equation 
[\BoghosianTaylor].  More complicated QLGA can be constructed to model 
potentials [\qlgaI], inhomogeneities and boundary conditions 
[\qlgaII].  In this talk we motivate the introduction of a QLGA model 
from a completely new perspective---Parrondo games.

%--------|---------|---------|---------|---------|---------|---------|
A Parrondo game is a sequence of plays of two simpler games, each of
which involves flipping biased coins.  In \S1 we review the somewhat 
surprising result that even if each of the simpler games is a losing
game, an alternating sequence of them can be a winning game 
[\Parrondo,\HarmerAbbott].  Meyer has recently initiated the study of
quantum game theory with an example of a coin flipping game, \PQpf\
[\qstrat].  This raises the natural question:  Is there a quantum 
version of Parrondo games?  Although the quantum Parrondo game we 
construct is not a two player game (as \PQpf\ is) it introduces a 
formalism for coherently iterated games which we expect to be useful 
in contexts involving one, two, or more players.

%--------|---------|---------|---------|---------|---------|---------|
Parrondo invented the coin flipping game, however, to illustrate a 
physical phenome\-non---Brownian ratchets [\Parrondo,\HarmerAbbott]; 
in \S2 we explain this connection in terms of a probabilistic 
discrete model---a random walk.  This stochastic microscopic model 
captures the macroscopic irreversible behavior of ratcheting, but
raises the concern that a microscopic quantum model which is exactly
unitary may not be able to do so [\Abbott].  The more immediate 
difficulty is the absence of any unitary version of a random walk.  To 
get a `quantizable' model we must first generalize to a 
{\sl correlated\/} random walk [\Taylor], or equivalently, a 
probabilistic LGA; we explain this in \S3.

%--------|---------|---------|---------|---------|---------|---------|
{}From here it is only a small step---actually an analytic 
continuation [\DWMF]---to a single particle QLGA.  We review the 
unitary evolution rules in \S4, emphasizing the inclusion of 
potentials which are necessary to model ratcheting.  \S5 contains the 
results of simulations which appear to illustrate quantum ratcheting,
and which lead us to answer our motivating question by interpreting 
the single particle QLGA with appropriate potentials as a quantum 
Parrondo game.  We conclude in \S6 with a summary and some more 
physical observations.

\medskip
\noindent{\bf 1.  Parrondo games}

%--------|---------|---------|---------|---------|---------|---------|
\noindent Consider games which involve flipping a coin:  winning 1 
when it lands head up and losing 1 when it lands tail up.  Suppose
there are three biased coins $A$, $B_0$, and $B_1$, with probabilities 
of landing head up of $p_a$, $p_0$, and $p_1$, respectively.  Define 
game $A$ to consist of repeatedly flipping coin $A$.  For 
$p_a < {1\over2}$, $A$ is a losing game in the sense that if the
initial stake is $x = 0$, after $t$ plays the expected value of the
payoff is $\langle x\rangle = t(2p_a - 1) < 0$.  Even though one may 
win sometimes, in the long run one must expect to lose.

%--------|---------|---------|---------|---------|---------|---------|
After each flip the payoff $x$ changes by $\pm1$.  Define game $B$ to
consist of repeatedly flipping coins $B_0$ and $B_1$:  $B_0$ when
$x \equiv 0$ (mod 3) and $B_1$ otherwise.  This defines a Markov 
process on $x$ (mod 3) with transition matrix
$$
T_B = \pmatrix{  0   & 1-p_1 &  p_1  \cr
                p_0  &   0   & 1-p_1 \cr
               1-p_0 &  p_1  &   0   \cr
              }.                                              \eqno(1)
$$
The equilibrium state, \ie, the eigenvector $(v_0,v_1,v_2)$ of $T_B$ 
with eigenvalue 1 (normalized by $v_i \ge 0$, $\sum v_i = 1$) 
determines the long time behavior of the game:  for large $t$, the
expected payoff is
$\langle x\rangle = t[(2p_0 - 1)v_0 + (2p_1 - 1)(v_1 + v_2)]$.  Thus
$B$ is a fair game iff the matrix
$$
\pmatrix{  -1   &  1-p_1 &   p_1  \cr
           p_0  &   -1   &  1-p_1 \cr
         2p_0-1 & 2p_1-1 & 2p_1-1 \cr
        }                                                     \eqno(2)
$$
is singular, \ie, iff
$$
p_0 = {1 - 2p_1 + p_1^2 \over 1 - 2p_1 + 2p_1^2}.             \eqno(3)
$$
One specific solution to equation (3) is 
$(p_0,p_1) = ({1\over10},{3\over4})$, but for a Parrondo game, $B$ 
should be a losing game, which means choosing $p_0$ and $p_1$ such 
that $\hbox{\eightpoint LHS}(3) < \hbox{\eightpoint RHS}(3)$.  
Figure~1 plots $\langle x\rangle$ as a function of $t$ for 
$A$ and $B$ games defined by $p_a = {1\over2} - \epsilon$, 
$p_0 = {1\over10} - \epsilon$,

\parskip=0pt
\parshape=1
0pt \halfwidth
\noindent  and $p_1 = {3\over4} - \epsilon$, with
$\epsilon = 0.005$.  Each is clearly a losing game.

\parskip=10pt
%--------|---------|---------|---------|---------|---------|---------|
\moveright\secondstart\vtop to 0pt{\hsize=\halfwidth
\vskip -3\baselineskip
$$
\epsfxsize=\halfwidth\epsfbox{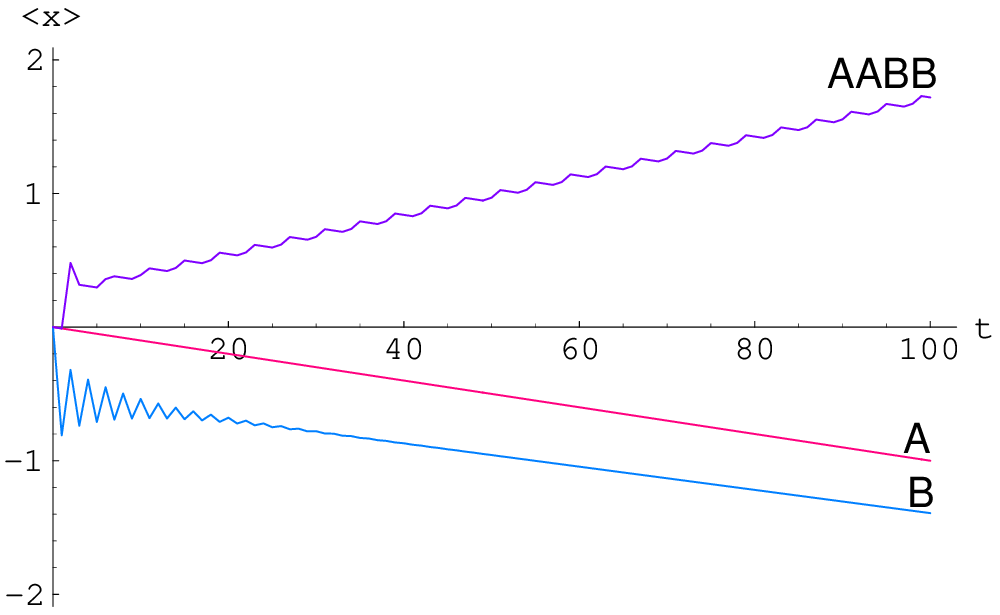}
$$
\vskip 0\baselineskip
\eightpoint{%
\noindent{\bf Figure~1}.  The expected payoffs for games $B$, $A$ and
$AABB$ as a function of number of plays $t$.  Although $A$ and $B$ are 
losing games, the combination $AABB$ is a winning game.
}}
\vskip -\baselineskip
\parshape=13
0pt \halfwidth
0pt \halfwidth
0pt \halfwidth
0pt \halfwidth
0pt \halfwidth
0pt \halfwidth
0pt \halfwidth
0pt \halfwidth
0pt \halfwidth
0pt \halfwidth
0pt \halfwidth
0pt \halfwidth
0pt \hsize
%--------|---------|---------|---------|---------|---------|---------|
Now suppose we combine these games.  More precisely, suppose they are
played in the order $AABB$, repeatedly.  Figure~1 plots the expected
result of this game as well.  Parrondo's `paradoxical' observation is
that this combination of two losing\break
games is a winning game!  To 
understand this phenomenon, rather than attempting to generalize the
Markov process analysis of equations (1)--(3), let us go back to the 
physical system which motivated Parrondo.

\medskip
\noindent{\bf 2.  Brownian ratchets}

%--------|---------|---------|---------|---------|---------|---------|
\noindent The payoff $x$ for game $A$ with $p_a = {1\over2}$ executes
an unbiased random walk on the integers, which is a discrete model for
the diffusion equation in $1+1$ dimensions [\diffusion]:
$$
\rho_t = D \rho_{xx}.                                         \eqno(4)
$$
That is, the distribution 
$p(x,t) = {\rm Prob}({\rm payoff} = x {\rm\ at\ time} = t)$ 
approximates $\rho(x,t)$ in (4) with $D = (\Delta x)^2/2\Delta t$.  
For $p_a \not= {1\over2}$ the random walk is biased and is a discrete
model for diffusion with linear advection [\diffusion]:
$$
\rho_t + c \rho_x = D \rho_{xx},                              \eqno(5)
$$
where $c = (2p_a - 1) \Delta x/\Delta t$.  Equation (5) describes
Brownian motion of a particle in a linear potential 
$V_A(x) \propto -(2p_a - 1)x$; the particle diffuses and tends 
downhill, as shown in Figure~2%
\sfootnote{$^*$}{Figures 2--4 correspond to the same {\sl exact\/}
calculation of the distributions of payoffs for which the expectation
values are plotted in Figure~1.  To compensate for the familiar 
$\sZ_2$ `spurious' conserved quantity in $1+1$ dimensional LGA
[\spurious], the `$t = 100$' distributions plotted in Figures~2--4 are 
actually $[p(x,99) + 2p(x,100) + p(x,101)]/4$.}
for  the case $p_a = {1\over2} -\epsilon$ simulated in \S1.

%--------|---------|---------|---------|---------|---------|---------|
\topinsert
\null\vskip 0\baselineskip
$$
\epsfxsize=\halfwidth\epsfbox{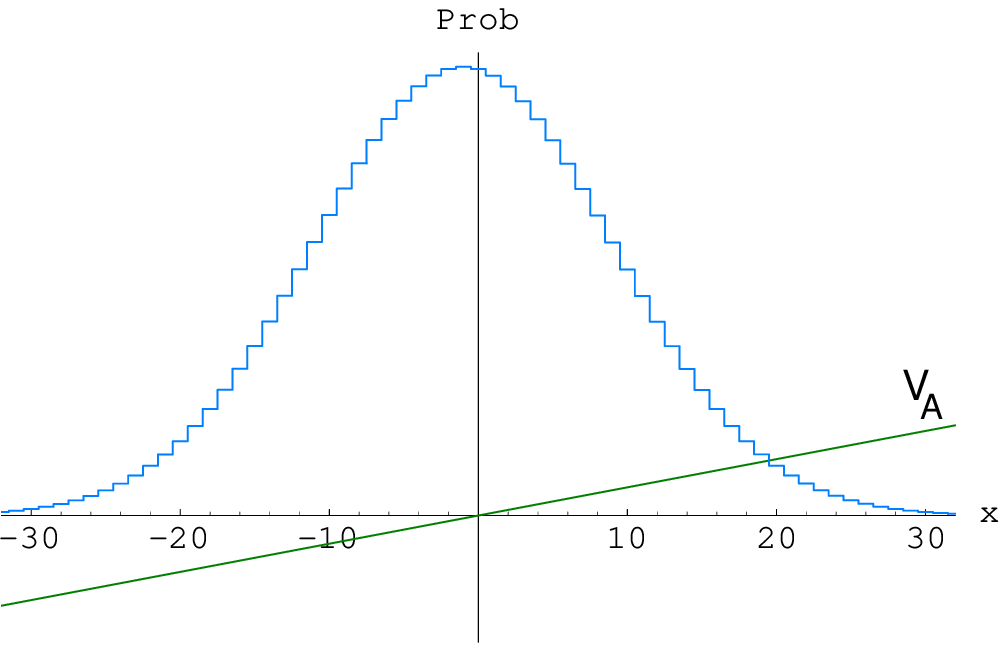}\hskip\chasm%
\epsfxsize=\halfwidth\epsfbox{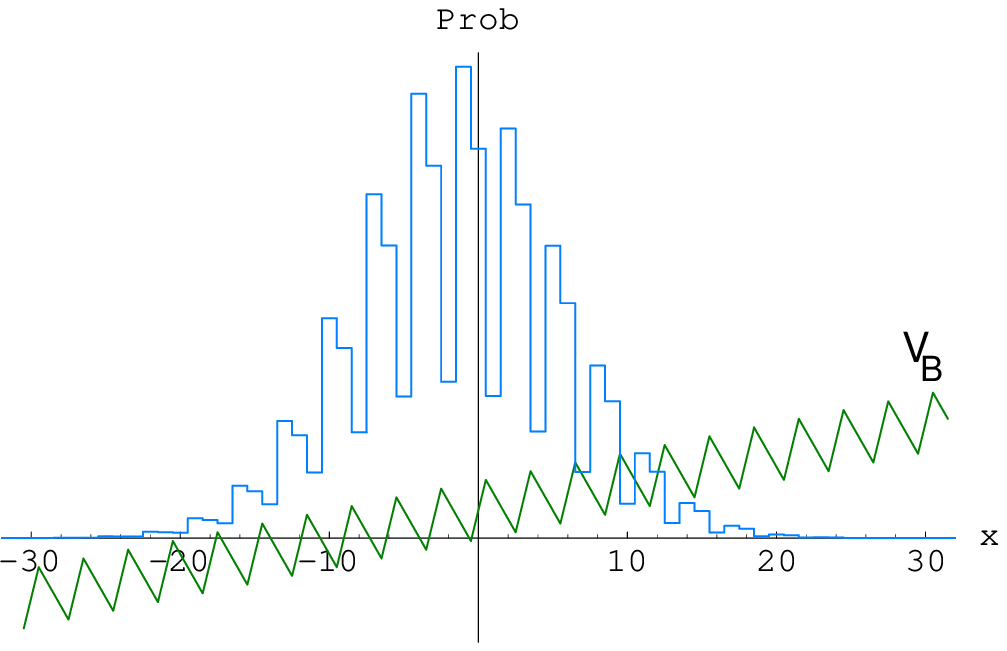}
$$
%\vskip-1.5\baselineskip
\hbox to\hsize{%
\vbox{\hsize=\halfwidth\eightpoint{%
\noindent{\bf Figure~2}.  The payoff distribution for game $A$ after
100 plays.  $V_A(x)$ is also graphed, in different vertical units.  
The initial distribution concentrated at $x=0$ has spread and shifted 
downhill; the peak is now at $-1$.
}}
\hfill%
\vbox{\hsize=\halfwidth\eightpoint{%
\noindent{\bf Figure~3}.  The payoff distribution for game $B$ after
100 plays.  $V_B(x)$ is also graphed, in different vertical units.  
The initial distribution concentrated at $x=0$ has spread and 
concentrated in the valleys of $V_B$, but also shifted downhill.
}}}
\endinsert

%--------|---------|---------|---------|---------|---------|---------|
Similarly, game $B$ corresponds to Brownian motion of a particle in a
piecewise linear potential.  For a fair game, \ie, for $p_0$ and $p_1$
satisfying equation (3), the potential (as well as its gradient) is 
periodic:
$$
V(x) \propto \cases{ -(2p_0 - 1)x & if $|x -3 n| \le b$, $n\in\Z$; \cr
                     -(2p_1 - 1)x & otherwise.                     \cr
                   }                                          \eqno(6)
$$
Here we assume $0 \le p_0 < {1\over2} < p_1 < \min\{1,(3-4p_0)/2\}$
and hence
$$
0 < b = {3(2p_1 - 1)\over 4(p_1 - p_0)} < 1
$$
makes the piecewise linear potential {\sl continuous}.  For the losing
game $B$ simulated in \S1, subtracting $\epsilon$ from the fair game
probabilities ${1\over10}$ and ${3\over4}$ for $p_0$ and $p_1$ 
corresponds to adding the $A$ game potential to the fair $B$ game 
potential of (6):  $V_B(x) = V(x) + V_A(x)$.  In this potential, as 
shown in Figure~3$^*$, the particle diffuses, concentrates in valleys, 
and tends downhill.
\vfill\eject

%--------|---------|---------|---------|---------|---------|---------|
\moveright\secondstart\vtop to 0pt{\hsize=\halfwidth
\vskip -1\baselineskip
$$
\epsfxsize=\halfwidth\epsfbox{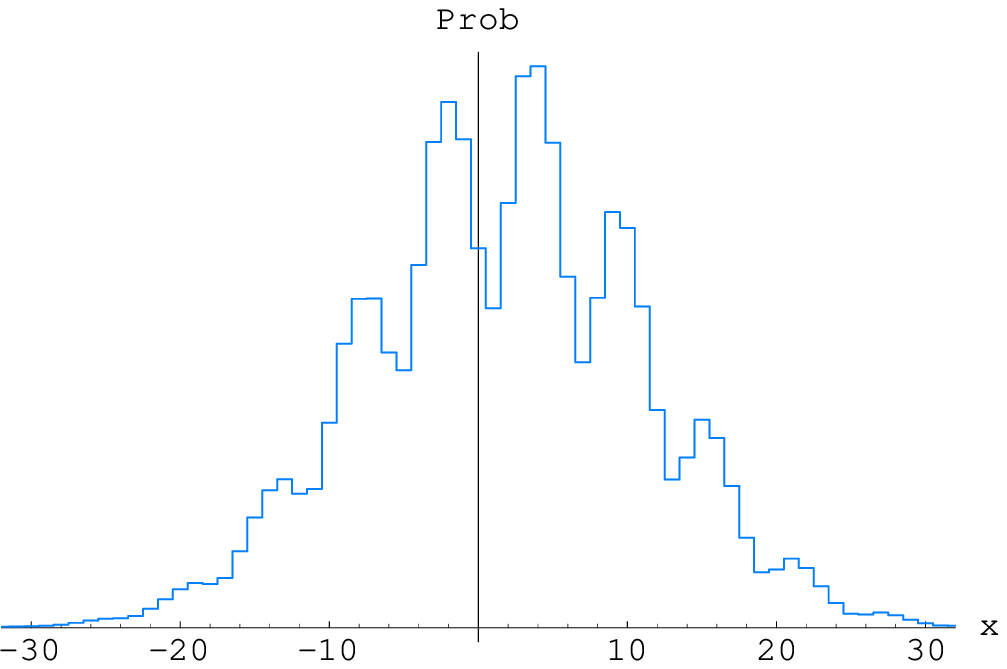}
$$
\vskip -0.5\baselineskip
\eightpoint{%
\noindent{\bf Figure~4}.  The payoff distribution for the alternating
game $AABB$ after 100 plays.  Although the initial distribution has 
spread and concentrated, it has shifted {\sl uphill}.
}}
\vskip -\baselineskip
\parshape=15
0pt \halfwidth
0pt \halfwidth
0pt \halfwidth
0pt \halfwidth
0pt \halfwidth
0pt \halfwidth
0pt \halfwidth
0pt \halfwidth
0pt \halfwidth
0pt \halfwidth
0pt \halfwidth
0pt \halfwidth
0pt \halfwidth
0pt \halfwidth
0pt \hsize
%--------|---------|---------|---------|---------|---------|---------|
Finally, Figure~4$^*$ shows the distribution of payoffs for the 
combined $AABB$ game.  Alternating the games models a `flashing'
potential [\flashing], which allows diffusion uphill during $A$ to be
concentrated into uphill valleys by $B$, leading to an average 
movement {\sl uphill}.  This phenomenon illustrates the use of a
ratchet as a thermal engine, first explained by Smoluchowski 
[\Smoluchowski] and subsequently discussed by Feynman [\Feynman], by 
Parrondo and Espa\~nol [\ParrondoEspanol], and by Abbott, Davis and
Parrondo [\ADP].  Such Brownian ratchets have been created 
experimentally in electromechanical [\electro] and optical [\optical] 
systems.

%--------|---------|---------|---------|---------|---------|---------|
Recognizing Parrondo games as Brownian ratchets raises concerns about
constructing quantum mechanical versions of them [\Abbott]:  the 
thermal ratchet engine works only for systems which are 
out-of-equilibrium (they require heat baths at two different
temperatures) and dissipative (the pawl must bounce inelastically off
the ratchet).  It is hard to imagine exactly unitary systems modelling
either of these properties.  In fact, recent theoretical analysis
[\qrattheory] and experimental observation [\qratexp] of quantum 
ratcheting have depended on some degree of dissipation/decoherence.  
Our goal, in contrast, is an exactly {\sl unitary\/} model.

\medskip
\noindent{\bf 3.  Correlated random walks}

%--------|---------|---------|---------|---------|---------|---------|
\noindent The first obstacle we must overcome is the non-existence of
a quantum random walk.  More precisely, there is no nontrivial unitary
band diagonal matrix which would describe the transition amplitudes
from each lattice site to some neighboring set of lattice sites 
[\nogo].  The intuition for this result is that the evolution of 
nontrivial classical random walks is not invertible and unitarity is
simply the quantum manifestation of invertibility.

%--------|---------|---------|---------|---------|---------|---------|
\moveright\secondstart\vtop to 0pt{\hsize=\halfwidth
\vskip 0\baselineskip
$$
\epsfxsize=\halfwidth\epsfbox{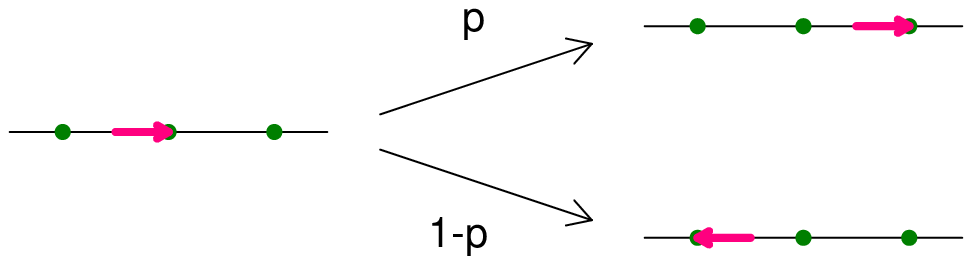}
$$
\vskip 0\baselineskip
\eightpoint{%
\noindent{\bf Figure~5}.  Evolution rules for a correlated random 
walk.  The reflected rules may or may not have the same probabilities;
if they do not, the random walk is biased.
}}
\vskip -\baselineskip
\parshape=12
0pt \halfwidth
0pt \halfwidth
0pt \halfwidth
0pt \halfwidth
0pt \halfwidth
0pt \halfwidth
0pt \halfwidth
0pt \halfwidth
0pt \halfwidth
0pt \halfwidth
0pt \halfwidth
0pt \hsize
%--------|---------|---------|---------|---------|---------|---------|
To construct an invertible model we must add an extra bit of 
information to each lattice site in $\Z$, the direction from which the
particle reached that site.  Figure~5 illustrates such a model:  the
arrows pointing to lattice sites record the direction from which the
site was reached and the (probabilistic) evolution rule shown is that 
the particle has probability $p$ of continuing in the same direction
and probability $1-p$ of changing direction.  This is a 
{\sl correlated\/} random walk [\Taylor]:  the probabilities for 
successive steps  are not independent for $p \not= {1\over2}$.  For
$p = {1\over2}$, however, they are uncorrelated, so this model 
specializes to the standard random walk.  To obtain an uncorrelated
but biased random walk, the probabilities should be independent of the
previous outcome, but not symmetric under reflection (\ie, parity).

%--------|---------|---------|---------|---------|---------|---------|
We can also think of this as a probabilistic LGA.  The extra bit of 
information is the particle momentum and, as we have described it, one 
timestep of the evolution consists of two parts:  scattering, defined 
by a stochastic matrix
$$
S = \bordermatrix{            & \leftarrow & \rightarrow \cr
                   \leftarrow &      p     &     1-p     \cr
                  \rightarrow &     1-p    &      p      \cr
                 },                                           \eqno(7)
$$
followed by advection.  Although this is the opposite order to the
usual way we think of LGA evolution, the two only differ by a time
translation of `half a timestep'.  In fact, long before the earliest
LGA were constructed to model fluid flow [\fluidflow], Goldstein 
[\Goldstein] and Kac [\Kac] showed that this probabilistic LGA is a
discrete model for a physical system---a $1+1$ dimensional wave 
equation with dissipation:
$$
{1\over v^2}\phi_{tt} + {2a\over v^2}\phi_t - \phi_{xx} = 0,
$$
where $v = \Delta x/\Delta t$ and $a = (1-p)/\Delta t$.  The $a \to 0$
limit of this `telegrapher equation' is the wave equation, and the
$a,v \to \infty$ limit with $v^2/2a = D$ is the diffusion equation 
(4).

%--------|---------|---------|---------|---------|---------|---------|
This correlated random walk/probabilistic LGA corresponds to a 
generalization of coin flipping games in which the probability of 
winning each play depends on the outcome of the previous play, and 
thus provides a framework in which to generalize Parrondo games.  
Parrondo, Harmer and Abbott have also introduced a generalization in 
which the probability of winning each play depends on the history of 
the game to that point---the past two outcomes in their 
case---although their motivation is to eliminate the $x$ dependence of 
the game [\PHA].  Our motivation is different:  we want to preserve 
this dependence, since it corresponds to a spatially varying 
potential, but use the generalization instead to construct unitary 
versions of these games.

\medskip
\noindent{\bf 4.  Quantum lattice gas automata}

%--------|---------|---------|---------|---------|---------|---------|
\noindent Now that we have a {\sl stochastic\/} scattering matrix (7),
it is straightforward to replace it with a {\sl unitary\/} matrix
$$
U = \bordermatrix{            &  \leftarrow & \rightarrow \cr
                   \leftarrow &  \cos\theta & i\sin\theta \cr
                  \rightarrow & i\sin\theta &  \cos\theta \cr
                 },                                           \eqno(8)
$$
although we must reinterpret the state space of the LGA to do so.  Let
$|x,\alpha\rangle$ denote the presence of a particle at lattice site
$x \in \Z$ with momentum $\alpha \in \{\pm1\}$.  States of the 
probabilistic LGA are convex combinations 
$$
f = \sum f_{x,\alpha} |x,\alpha\rangle,
\quad {\rm with\ } 0 \le f_{x,\alpha} \in \R
{\rm\ and\ }\sum f_{x,\alpha} = 1,                           \eqno(9)
$$ 
so that $f_{x,\alpha}$ is the probability that the particle is in the 
state $|x,\alpha\rangle$.  Evolution consists of scattering according
to (7):
$$
\eqalignno{
|x,\alpha\rangle 
 &\mapsto p|x,\alpha\rangle + (1-p)|x,-\alpha\rangle,              \cr
\noalign{\hbox{followed by advection}}
 &\mapsto p|x+\alpha,\alpha\rangle + (1-p)|x-\alpha,-\alpha\rangle,\cr
}
$$
extended by linearity to general states $f$ (9).

%--------|---------|---------|---------|---------|---------|---------|
For a QLGA, the general one particle state is a vector in Hilbert 
space [\qcaqlg,\qlgaI,\qlgaII,\lgbu]:
$$
\psi = \sum \psi_{x,\alpha} |x,\alpha\rangle,
\quad {\rm with\ } \psi_{x,\alpha} \in \C
{\rm\ and\ }\sum |\psi_{x,\alpha}|^2 = 1,                    \eqno(10)
$$
so that $\psi_{x,\alpha}$ is the {\sl amplitude\/} of the state
$|x,\alpha\rangle$ and $|\psi_{x,\alpha}|^2$ is the probability that,
if measured in this basis, the particle is observed to be in state
$|x,\alpha\rangle$.  Quantum evolution consists of scattering 
according to (8):
$$
\eqalignno{
|x,\alpha\rangle 
 &\mapsto \cos\theta|x,\alpha\rangle 
       + i\sin\theta|x,-\alpha\rangle,                             \cr
\noalign{\hbox{followed by advection}}
 &\mapsto \cos\theta|x+\alpha,\alpha\rangle 
       + i\sin\theta|x-\alpha,-\alpha\rangle,                      \cr
}
$$
extended by linearity to general states $\psi$ (10).  This evolution 
is unitary because the scattering stage is, and because the advection 
is deterministic.  Furthermore, we can include a potential with 
multiplication by an $x$-dependent phase $e^{-iV(x)}$ [\qlgaI,\lgbu]; 
the evolution remains unitary.  The problem thus reduces to picking
parameters $\theta$, $V(x,t)$ to achieve ratcheting---which we can 
also interpret as a quantum Parrondo phenomenon.

\medskip
\noindent{\bf 5.  Quantum Parrondo games}

%--------|---------|---------|---------|---------|---------|---------|
\noindent Since we are going to exhibit our results as outputs of
simulations, we should first remark that although we may think of our
single particle QLGA as a particle moving from lattice site to lattice
site with specified amplitudes, on a classical computer we must
simulate it using a lattice Boltzmann method.  That is, we must keep
track of the whole vector $\psi$ and evolve that at each timestep.  In
fact, this is how we performed the exact computations for the 
probabilistic LGA for Figures~1--4.  In the probabilistic case we have 
the option of simulating it as a lattice gas and averaging over 
multiple runs---the results of Harmer and Abbott were obtained this 
way, using 50,000 runs [\HarmerAbbott]---but for the quantum case we 
do not have this option.

%--------|---------|---------|---------|---------|---------|---------|
\topinsert
\null\vskip -3\baselineskip
$$
\epsfxsize=\halfwidth\epsfbox{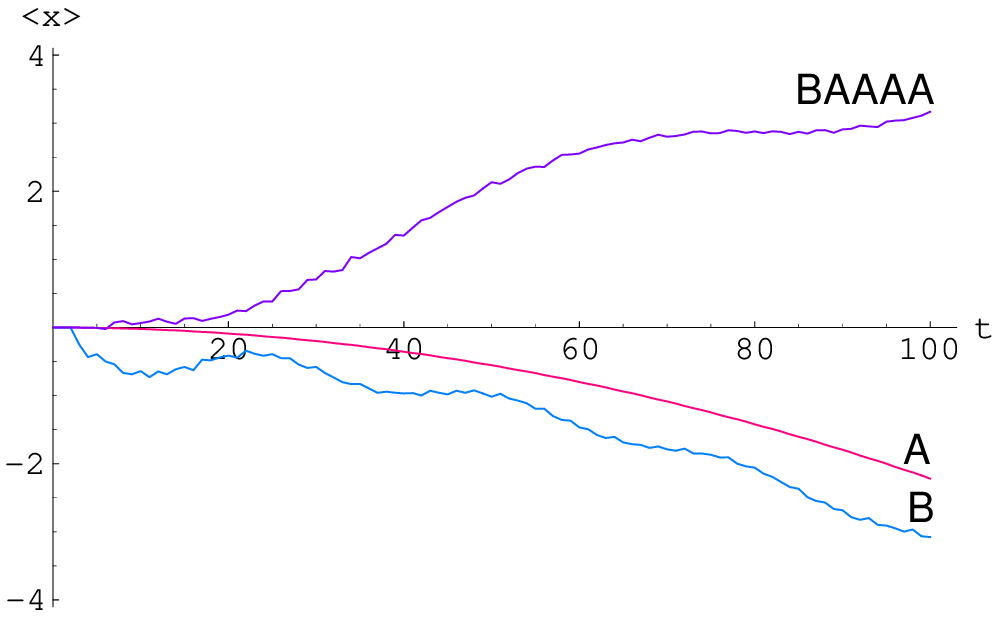}\hskip\chasm%
\epsfxsize=\halfwidth\epsfbox{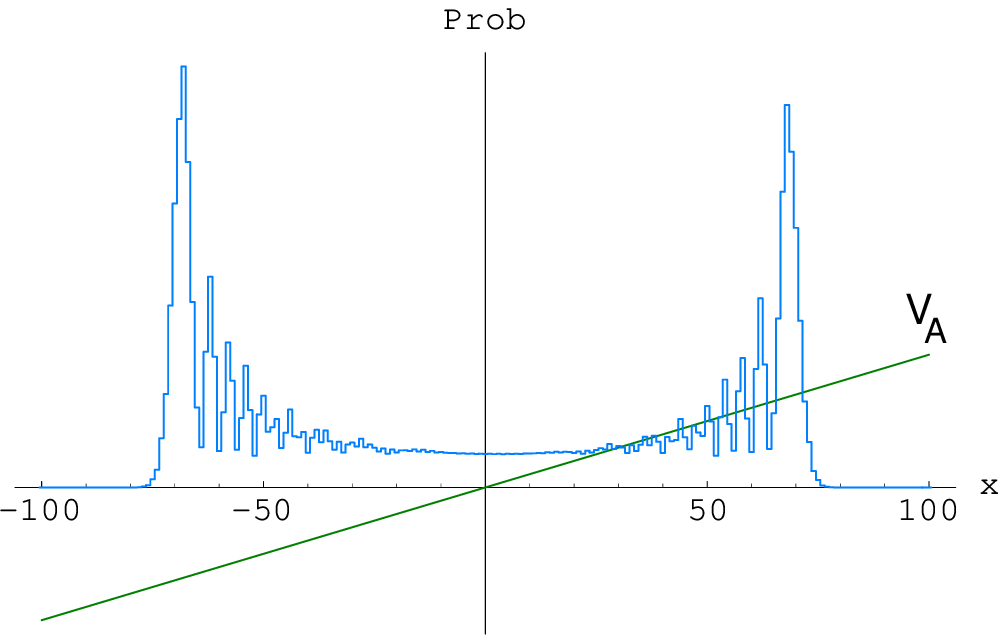}
$$
%\vskip-1.5\baselineskip
\hbox to\hsize{%
\vbox{\hsize=\halfwidth\eightpoint{%
\noindent{\bf Figure~6}.  The expected payoffs for quantum games $B$, 
$A$ and $BAAAA$ as a function of number of plays $t$.  Although $A$ 
and $B$ are losing games, the combination $BAAAA$, played repeatedly,
is a winning game over this range of numbers of plays.  These results 
illustrate the same `paradoxical' phenomenon as those in Figure~1 do 
for the classical Parrondo game.
}}
\hfill%
\vbox{\hsize=\halfwidth\eightpoint{%
\noindent{\bf Figure~7}.  The payoff distribution for quantum game $A$ 
after 100 plays.  $V_A(x)$ is also graphed, in different vertical 
units.  The initial distribution concentrated at $x=0$ contained equal 
left and right moving amplitudes which have shifted to peaks at about 
$\pm68$ and spread.  Interference has created a series of peaks at 
smaller absolute payoffs; the average has shifted slightly downhill.
}}}
\endinsert

%--------|---------|---------|---------|---------|---------|---------|
We set $\theta = {\pi\over4}$ in (8) so that the magnitudes of the 
amplitudes are all the same---this is the analogue of an unbiased,
uncorrelated random walk.  The initial state is an equal superposition 
of $|0,-1\rangle$ and $|0,+1\rangle$ so that there is the same initial
capital---zero---as in the classical simulations, and no bias for the
initial momentum/state at $t = -1$.  Figure~6 shows the expectation
value $\langle x \rangle$ as a function of $t$ for
$$
V_A(x) = {2\pi\over5000} x \quad{\rm and}\quad
V_B(x) = {\pi\over3} [1 - {1\over2}(x {\rm\ mod\ }3)] + V_A(x).
$$
As in the classical case, $V_A$ is a linear potential, as shown in 
Figure~7, and $V_B$ is a piecewise linear 3-periodic potential 
superimposed on $V_A$, as shown in Figure~8.  Figure~6 shows that the
behavior is similar to the classical cases:  Potentials $V_A$ and 
$V_B$ individually force $\langle x \rangle$ downhill, but the 
flashing pattern---$BAAAA$, repeated---drives $\langle x \rangle$
{\sl uphill}.  (We chose the parameters in these potentials to produce
expectation value curves similar to those shown in the classical 
cases; they differ by only about a factor of 2 after 100 plays.)  As 
shown in Figure~7, the evolution in $V_A$ is biased downhill, but 
looks very little like the diffusive evolution of Figure~2.  
Similarly, as shown in Figure~8, the evolution in $V_B$ is biased 
downhill, and concentrates periodically, but otherwise looks quite 
different than the\break

%--------|---------|---------|---------|---------|---------|---------|
\midinsert
\null\vskip -3\baselineskip
$$
\epsfxsize=\halfwidth\epsfbox{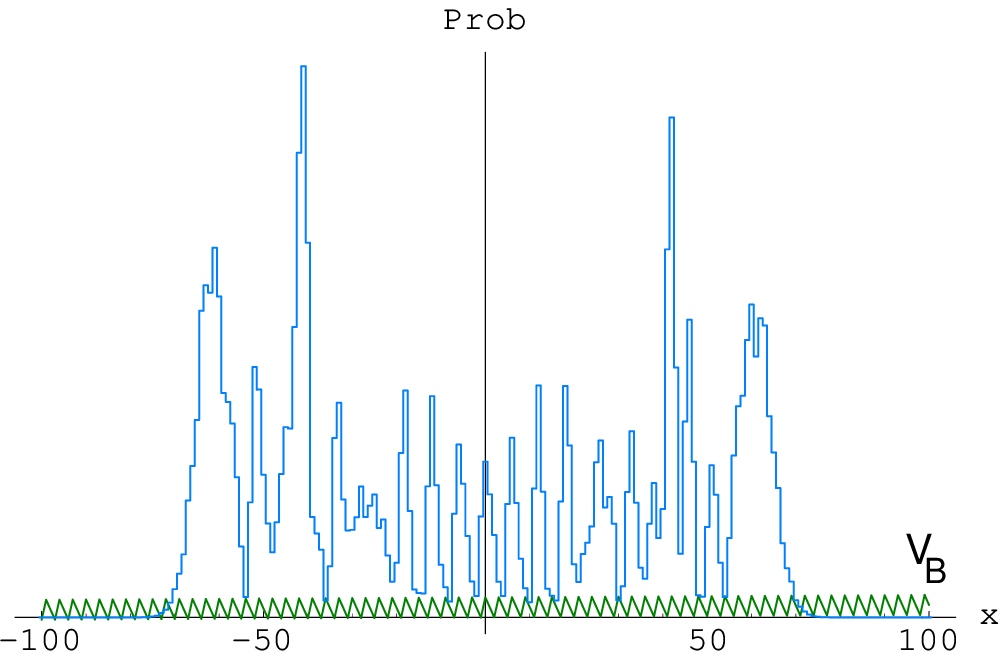}\hskip\chasm%
\epsfxsize=\halfwidth\epsfbox{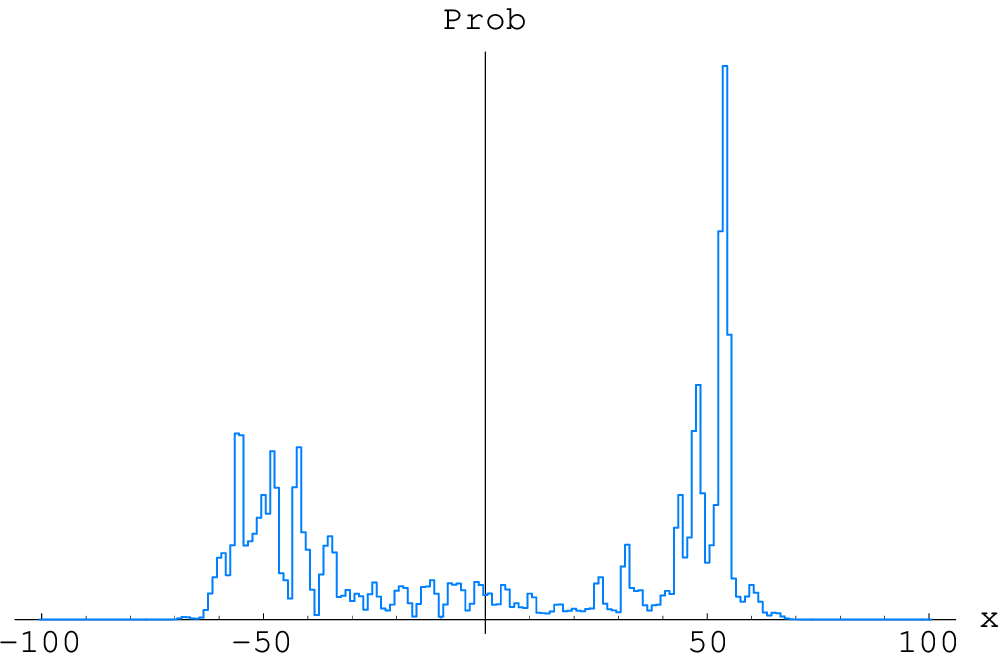}
$$
%\vskip-1.5\baselineskip
\hbox to\hsize{%
\vbox{\hsize=\halfwidth\eightpoint{%
\noindent{\bf Figure~8}.  The payoff distribution for quantum game $B$ 
after 100 plays.  $V_B(x)$ is also graphed, in different vertical 
units.  The initial distribution has shifted left/right, and 
spread.  $V_B$ has caused a more complicated interference pattern than 
$V_A$, but the average has also shifted downhill.
}}
\hfill%
\vbox{\hsize=\halfwidth\eightpoint{%
\noindent{\bf Figure~9}.  The payoff distribution for the alternating
game $BAAAA$ after 100 plays.  The distribution still shows the 
results of interference, but the large positive peak slightly 
outweighs the large negative peak to give an average which has moved 
{\sl uphill}.
}}}
\endinsert

\vfill\eject

\noindent diffusive evolution of Figure~3.  Finally, as shown 
in Figure~9, flashing the potentials in the order $BAAAA$, repeatedly, 
biases the evolution uphill, but still in a way unlike the classical 
case of Figure~4.  Interpreting this QLGA as a quantum Parrondo game, 
Figure~9 shows that this is a game for gamblers with high tolerance 
for risk---the large probability of a big loss is just barely 
outweighed by the slightly larger probability of a big win.

\medskip
\noindent{\bf 6.  Conclusions}

%--------|---------|---------|---------|---------|---------|---------|
\noindent By interpreting classical Parrondo games as probabilistic 
LGA, we have motivated the introduction of QLGA to answer the 
question:  Are there quantum Parrondo games?  The simulations shown in 
\S5 appear to answer this question in the affirmative, as well as to 
demonstrate discrete quantum ratcheting, despite the absence of 
dissipation.

%--------|---------|---------|---------|---------|---------|---------|
\moveright\secondstart\vtop to 0pt{\hsize=\halfwidth
\vskip -1\baselineskip
$$
\epsfxsize=\halfwidth\epsfbox{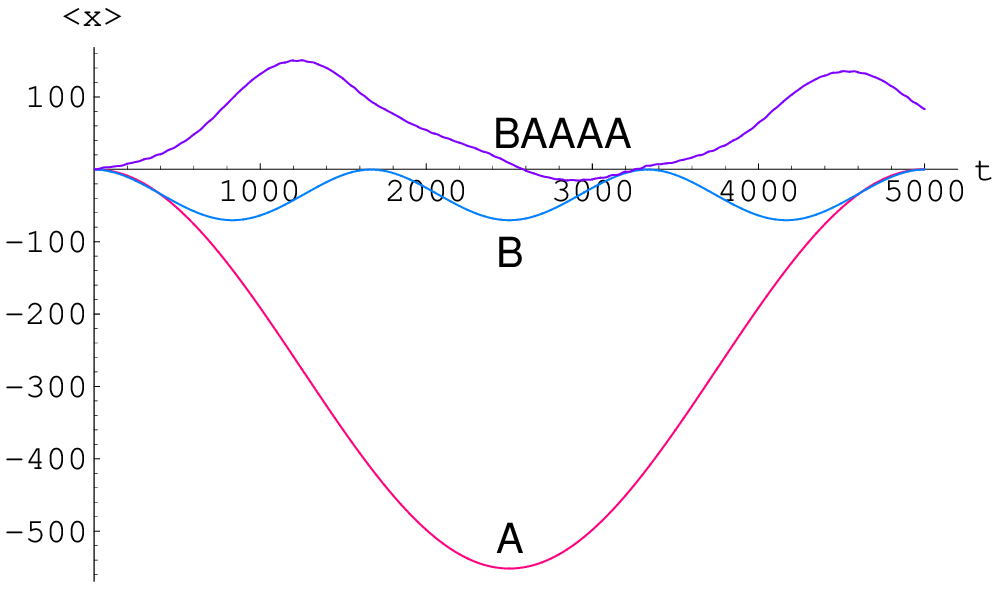}
$$
\vskip 0\baselineskip
\eightpoint{%
\noindent{\bf Figure~10}.  The expected payoffs for quantum games $A$, 
$B$ and $BAAAA$ as a function of number of plays $t$.  Although the 
curves are periodic, for random times (or on average), $A$ and $B$ 
have negative expected payoffs while $BAAAA$ has a positive expected 
payoff.
}}
\vskip -\baselineskip
\parshape=17
0pt \halfwidth
0pt \halfwidth
0pt \halfwidth
0pt \halfwidth
0pt \halfwidth
0pt \halfwidth
0pt \halfwidth
0pt \halfwidth
0pt \halfwidth
0pt \halfwidth
0pt \halfwidth
0pt \halfwidth
0pt \halfwidth
0pt \halfwidth
0pt \halfwidth
0pt \halfwidth
0pt \hsize
%--------|---------|---------|---------|---------|---------|---------|
The quadratic growths of $\langle x\rangle$ shown in Figure~6, 
however, should be worrisome since the single particle QLGA 
discretizes the Dirac equation [\qcaqlg,\qlgaI,\lgbu], which is 
relativistic.  If $\langle x\rangle$ were to continue to grow 
quadratically, it would eventually exit the lightcone---not 
relativistic behavior.  Figure~10 shows the results of simulation out
to $t = 5000$.  We see that the expectation values do not continue to 
grow quadratically; rather their evolution is oscillatory and the 
small $t$ quadratic growth is that of $A(\cos(b t) - 1)$.  In fact, 
the QLGA with potential $V_A$ discretizes the `Dirac oscillator' [\DO] 
which can be solved exactly, and in which wave packets are known to 
evolve approximately periodically [\NTC].  For random stopping 
times---\ie, measurement times---however, both the $A$ and $B$ games 
are losing games and the $BAAAA$ quantum game is a winning game.  In 
this sense the QLGA is a quantum Parrondo game.  In the broader 
context of game theory, it also illustrates a coherently repeated 
quantum game---and the possibility of interference between sequences 
of plays.  This kind of construction should generalize to, for 
example, a quantum version [\BenjaminHayden] of the 
M{\eightpoint INORITY} game [\minority].

%--------|---------|---------|---------|---------|---------|---------|
More physically, Aharonov, Ambainis, Kempe and Vazirani also use 
random stopping times to obtain a related result:  quantum (unitary) 
simulation of sampling from equilibrium distributions of diffusion 
processes on graphs with constant vertex degree [\AAKV].  The 
quadratic speedup they find is due to the linear in time (rather than 
$\sqrt{t}$ as in the classical random walk) spread of the wave 
function illustrated in Figure~7 [\NayakVishwanath].  More
generally, Childs, Farhi and Gutmann demonstrate the same quadratic
speedup for a {\sl continuous\/} time quantum process on certain
graphs, without the constant vertex degree restriction [\CFG].  Our 
results, and these, provide specific answers to the general question 
of whether quantum computers (see, \eg, [\NielsenChuang]) can 
calculate properties of classical systems more efficiently than can 
classical computers [\pqc].

\medskip
\noindent{\bf Acknowledgements}
\nobreak

\nobreak
%--------|---------|---------|---------|---------|---------|---------|
\noindent We thank Derek Abbott, Jeff Rabin and Ruth Williams for 
useful discussions.  This work has been partially supported by the 
National Security Agency (NSA) and Advanced Research and Development 
Activity (ARDA) under Army Research Office (ARO) contract number 
DAAG55-98-1-0376, and by the Air Force Office of Scientific Research
(AFOSR) under grant number F49620-01-1-0494.

\medskip
\global\setbox1=\hbox{[00]\enspace}
\parindent=\wd1

\noindent{\bf References}
\bigskip

\parskip=0pt
\item{[\FeynmanHibbs]}
R. P. Feynman and A. R. Hibbs,
{\sl Quantum Mechanics and Path Integrals\/}
(New York:  McGraw-Hill 1965).

\item{[\qcaqlg]}
\dajm,
``From quantum cellular automata to quantum lattice gases'',
\JSP\ {\bf 85} (1996) 551--574.

\item{[\BoghosianTaylor]}
B. M. Boghosian and W. Taylor, IV,
``A quantum lattice-gas model for the many-particle Schr\"odinger
  equation in $d$ dimensions'',
\PRE\ {\bf 8} (1997) 705--716.

\item{[\qlgaI]}
\dajm,
``Quantum mechanics of lattice gas automata:  one particle plane waves 
  and potentials'',
\PRE\ {\bf 55} (1997) 5261--5269.

\item{[\qlgaII]}
\dajm,
``Quantum mechanics of lattice gas automata:  boundary conditions and
  other inhomogeneities'',
\JPA\ {\bf 31} (1998) 2321--2340.

\item{[\Parrondo]}
J. M. R. Parrondo,
``Parrondo's paradoxical games'',
{\tt http://seneca.fis.ucm.es/ parr/GAMES}.

\item{[\HarmerAbbott]}
\ha,
``Parrondo's paradox'',
\StSc\ {\bf 14} (1999) 206--213;\hfb
\ha,
``Losing strategies can win by Parrondo's paradox'',
\Na\ {\bf 402} (1999) 864.

\item{[\qstrat]}
\dajm,
``Quantum strategies'',
\PRL\ {\bf 82} (1999) 1052--1055.

\item{[\Abbott]}
D. Abbott,
personal communication (February 2000).

\item{[\Taylor]}
G. I. Taylor,
``Diffusion by continuous movements'',
\PLMS\ {\bf 20} (1920) 196--212.

\item{[\DWMF]}
C. DeWitt-Morette and S. K. Foong,
``Path-integral solutions of wave equations with dissipation'',
\PRL\ {\bf 62} (1989) 2201--2204.

\item{[\diffusion]}
J. W. Strutt, Baron Rayleigh,
{\sl The Theory of Sound}, vol.\ 1, second edition
(London:  Macmillan 1894) \S42;\hfb
J. W. Strutt, Baron Rayleigh,
``On James Bernoulli's theorem in probabilities'',
\PM\ (5) {\bf 47} (1899) 246--251;\hfb
M. von Smoluchowski,
``{\it \"Uber Brownschen Molekularbewegung unter Einwirkung 
  \"au{\ss}er\-er Kr\"afte unde deren Zusammenhang mit der 
  verallgemeinerten Diffusionsgleichung}'',
\AdP\ {\bf 48} (1915) 1103--1112.

\item{[\spurious]}
G. Zanetti,
``Hydrodynamics of lattice gas automata'',
\PRA\ {\bf 40} (1989) 1539--1548;\hfb
Z. Cheng, J. L. Lebowitz and E. R. Speer,
``Microscopic shock structure in model particle systems:  The
  Boghosian-Levermore cellular automaton revisited'',
\CPAM\ {\bf XLIV} (1991) 971--979;\hfb
\bd,
``Lattice gases and exactly solvable models'',
\JSP\ {\bf 68} (1992) 575--590.

\item{[\flashing]}
C. R. Doering,
``Randomly rattled ratchets'',
\INCD\ {\bf 17} (1995) 685--697.

\item{[\Smoluchowski]}
M. von Smoluchowski,
``{\it Experimentell nachweisbare, der \"ublichen Thermodynamik 
       widersprechende Molekularph\"anomene}'',
\PZ\ {\bf XIII} (1912) 1069--1080.

\item{[\Feynman]}
R. P. Feynman, R. B. Leighton and M. Sands,
{\sl The Feynman Lectures on Physics}, vol.\ 1
(Reading, MA:  Addison-Wesley 1963), Sec.\ 46.1--46.9.

\item{[\ParrondoEspanol]}
J. M. R. Parrondo and P. Espa\~nol,
``Criticism of Feynman's analysis of the ratchet as an engine'',
\AJP\ {\bf 64} (1996) 1125--1130.

\item{[\ADP]}
D. Abbott, B. David and J. M. R. Parrondo,
``The problem of detailed balance for the Feynman-Smoluchowski engine
  (FSE) and the multiple pawl paradox'',
{\sl Unsolved Problems of Noise and Fluctuations, UPoN'99}, 
Proceedings of the Second International Conference, 
Adelaide, Australia, 12--15 July 1999, 
AIP Conference Proceedings {\bf 511} (2000) 213--218.

\item{[\electro]}
J. Rousselet, L. Salome, A. Adjari and J. Prost,
``Directional motion of Brownian particles induced by a periodic
  asymmetric potential'',
\Na\ {\bf 370} (1994) 446--448.

\item{[\optical]}
L. P. Faucheux, L. S. Bourdieu, P. D. Kaplan and A. J. Libchaber,
``Optical thermal ratchet'',
\PRL\ {\bf 74} (1995) 1504--1507.

\item{[\qrattheory]}
P. Riemann, M. Grifoni and P. H\"anggi,
``Quantum ratchets'',
\PRL\ {\bf 79} (1997) 10--13;\hfb
S. Yukawa, M. Kikuchi, G. Tatara and H. Matsukawa,
``Quantum ratchets'',
\JPSJ\ {\bf 66} (1997) 2953--2956;\hfb
G. Tatara, M. Kikuchi, S. Yukawa and H. Matsukawa,
``Dissipation enhanced-asymmet\-ric transport in quantum ratchets'',
\JPSJ\ {\bf 67} (1998) 1090--1093;\hfb
I. Goychuk, M. Grifoni and P. H\"anggi,
``Nonadiabatic quantum Brownian rectifiers'',
\PRL\ {\bf 81} (1998) 649--652.

\item{[\qratexp]}
A. Lorke, S. Wimmer, B. Jager, J. P. Kotthaus, W. Wegscheider and
M. Bichler,
``Far-infrared and transport properties of antidot arrays with 
  broken symmetry'',
\PB\ {\bf 249}--{\bf 251} (1998) 312-316;\hfb
H. Linke, T. E. Humphrey, A. L\"ofgren, A. O, Sushkov, R. Newbury,
R. P. Taylor and P. Omling,
``Experimental tunneling ratchets'',
\Sc\ {\bf 286} (1999) 2314--2317.

\item{[\nogo]}
\dajm,
``On the absence of homogeneous scalar unitary cellular automata'',
\PLA\ {\bf 223} (1996) 337--340.

\item{[\fluidflow]}
\hpdp,
``Time evolution of a two-dimensional model system.  I.  Invariant
  states and time correlation functions'',
\JMP\ {\bf 14} (1973) 1746--1759;\hfb
\hdpp,
``Molecular dynamics of a classical lattice gas:  transport 
  properties and time correlation functions'',
\PRA\ {\bf 13} (1976) 1949--1961;\hfb
U. Frisch, B. Hasslacher and Y. Pomeau,
``Lattice-gas automata for the Navier-Stokes equation'',
\PRL\ {\bf 56} (1986) 1505--1508.

\item{[\Goldstein]}
S. Goldstein,
``On diffusion by discontinuous movements, and on the telegraph 
  equation'',
\QJMAM\ {\bf 4} (1951) 129--156.

\item{[\Kac]}
M. Kac,
``A stochastic model related to the telegrapher's equation'',
\RMJM\ {\bf 4} (1974) 497--509;
reprinted from 
{\sl Some stochastic problems in physics and mathematics},
Colloquium Lectures in the Pure and Applied Sciences, No.~2 
(Dallas, TX:  Magnolia Petroleum and Socony Mobil Oil 1956).

\item{[\PHA]}
J. M. R. Parrondo, G. P. Harmer and D. Abbott,
``New paradoxical games based on Brownian ratchets'',
\PRL\ {\bf 85} (2000) 5226--5229.

\item{[\lgbu]}
\dajm,
``Quantum lattice gases and their invariants'',
\IJMPC\ {\bf 8} (1997) 717--735.

\item{[\DO]}
F. Sauter,
``{\it \"Uber das Verhalten eines Elektrons im homogenen elektrischen
  Feld nach der relativistischen Theorie Diracs\/}'',
\ZP\ {\bf 69} (1931) 742--764;\hfb
D. \^Ito, K. Mori and E. Carriere,
``An example of dynamical systems with linear trajectory'',
\NCA\ {\bf 51} (1967) 1119--1121;\hfb
P. A. Cook,
``Relativisitic harmonic oscillators with intrinsic spin structure'',
\LNC\ {\bf 1} (1971) 419--426;\hfb
M. Moshinsky and A. Szczepaniak,
``The Dirac oscillator'',
\JPA\ {\bf 22} (1989) L817--L819.

\item{[\NTC]}
Y. Nogami and F. M. Toyama,
``Coherent states of the Dirac oscillator'',
\CJP\ {\bf 74} (1996) 114--121;\hfb
F. M. Toyama, Y. Nogami and F. A. B. Continho,
``Behaviour of wavepackets of the `Dirac oscillator':  Dirac 
  representation versus Foldy-Wouthuysen representation'',
\JPA\ {\bf 30} (1997) 2585--2595.

\item{[\BenjaminHayden]}
S. C. Benjamin and P. M. Hayden,
``Multi-player quantum games'',
%{\tt quant-ph/0007038};
\PRA\ {\bf 64} (2001) 030301(R).

\item{[\minority]}
W. B. Arthur,
``Inductive reasoning and bounded rationality'',
\AER\ {\bf 84} (1994) 406--411;\hfb
D. Challet and Y.-C. Zhang,
``Emergence of cooperation and organization in an evolutionary
  game'',
\PA\ {\bf 246} (1997) 407--418.

\item{[\AAKV]}
D. Aharonov, A. Ambainis, J. Kempe and U. Vazirani,
``Quantum walks on graphs'',
{\tt quant-ph/0012090}.

\item{[\NayakVishwanath]}
A. Nayak and A. Vishwanath,
``Quantum walk on the line'',
{\tt quant-ph/0010117}.

\item{[\CFG]}
A. M. Childs, E. Farhi and S. Gutmann,
``An example of the difference between quantum and classical random
  walks'',
{\tt quant-ph/0103020}.

\item{[\NielsenChuang]}
M. A. Nielsen and I. L. Chuang,
{\sl Quantum Computation and Quantum Information\/}
(New York:  Cambridge University Press 2000).

\item{[\pqc]}
\dajm,
``Physical quantum algorithms'',
UCSD preprint (2001).

\bye